\documentclass{article}
\usepackage[accepted]{vietnam} 
\usepackage{natbib}
\usepackage{graphicx}      
\usepackage{wrapfig}
\begin{document}
\twocolumn[
\title{Astrochemistry Focus and Research on Planck Cold Clumps}
\titlerunning{Astrochemistry Focus and Research on Planck Cold Clumps}
\author{ KEN'ICHI TATEMATSU }{k.tatematsu@nao.ac.jp}
\address{Nobeyama Radio Observatory, National Astronomical Observatory of Japan,
National Institutes of Natural Sciences,
462-2 Nobeyama, Minamimaki, Minamisaku, Nagano 384-1305, Japan, and
Department of Astronomical Science,
SOKENDAI (The Graduate University for Advanced Studies), Japan}

\vskip 0.5cm 
]

\begin{abstract}

At the Astrochemistry Focus Group, we discussed
what is still missing in our understanding even with new knowledge given at this conference, and 
what can be done for that within 10 years from now.
Still missing in understanding are 
UV-photons and cosmic-rays interactions with icy dust grains,
Sulphur and Phosphorus chemistry,
Metallicity effect,
Duration (time) effect,
COM formation and destruction,
phase transition,
dust-gas interface,
dust evolution, etc.
What we should do are
multi-scale high spectral resolution molecular observations, laboratory work, theory, radiative transfer, etc.
We need careful modeling without simplifying things.
Next, I introduce our research on Planck cold clumps.
We observed thirteen Planck cold clumps with the James Clerk Maxwell Telescope/SCUBA-2 and with the Nobeyama 45 m radio telescope.  
The N$_2$H$^+$ distribution obtained with the Nobeyama telescope is quite similar to SCUBA-2 dust distribution.  The 82 GHz HC$_3$N, 82 GHz CCS, and 94 GHz CCS emission are often distributed  differently with respect to the N$_2$H$^+$ emission.  The CCS emission, which is known to be abundant in starless molecular cloud cores, is often very clumpy in the observed targets.  
We made deep single-pointing observations in DNC, HN$^{13}$C, N$_2$D$^+$, cyclic C$_3$H$_2$
toward nine clumps.  The detection rate of N$_2$D$^+$ is 50\%.  
In two of the starless clumps observe, the CCS emission is distributed as it surrounds the
N$_2$H$^+$ core (chemically evolved gas), which resembles the case of L1544, a prestellar core showing collapse. 
In addition, we detected both DNC and  N$_2$D$^+$.  These two clumps are most likely on the verge of star formation. 
We introduce the Chemical Evolution Factor (CEF) for starless cores to describe the chemical evolutionary stage, and analyze the observed Planck cold clumps.
\end{abstract}

\section{Summary of Discussion at the Astrochemistry Focus Group} 

\subsection{Goal}
Astrochemistry Focus Group wished to make clear what is still missing in our understanding even with new knowledge given at this conference.
The purpose is to have an idea on what can be done for that, say, within 10 years from now.

\subsection{Individual Comments}
Paola Caselli said 
``still missing things in the field of astrochemistry are good understanding of the dust-gas interface, dust evolution, UV-photons and cosmic-rays interactions with icy dust grains. 
Multi-scale high spectral resolution molecular observations, laboratory work, theory, and radiative transfer are needed.''
Alvaro Sanchez-Monge said 
``Carbon, Oxygen, Nitrogen chemistry is well understood... the next elements for which the chemistry still needs improvement will (probably) be Sulphur and Phosphorus.''

Next we discussed the environmental effect.
Nami Sakai said 
``Chemical diversity in YSO would originate from differences in time after the UV shielding of the molecular cloud (i.e. period in starless phase in each source).  Then, what can make the time difference?''
Yuri Nishimura said
``We need to be more aware of the effect of metallicity, that is, the amount of dust grains and UV-shielding.''
Takashi Shimonishi said
``Comprehensive understanding of gas-grain chemistry as a function of galactic metallicities is needed.
Statistical observations of ices and molecular gas around high- and low-mass YSOs in Local Group galaxies (LMC, SMC, IC10, M51, etc.) are important.
Also, I am interested in astrochemistry as a function of redshift.''

Kotomi Taniguchi pointed out that the
chemical evolution and mechanisms in high-mass star forming regions should be explored.
Kuo-Song Wang  said that he wants to know
``How complex molecules, especially those closely related to astrobiology, form, survive, and destroy during the star/planet formation processes?''
Alvaro Sanchez-Monge is concerned 
``How COMs are produced? Grain-surface vs gas-phase reactions. Are COMs expected to be found everywhere? (cold vs hot environments, low-mass vs high-mass dense cores, is the non-detection of COMs just a matter of sensitivity)?''
Gwendoline St\'{e}phan pointed out that
a better understanding of the formation routes of complex chemistry like hydrocarbons and COMs is required, whatever the regions. 
Nguyen Hoang Phuong Thanh is concerned
how to determine clearly about the chemical origin of CO and N2 differential depletion in prestellar core.
Yoko Okada asked
``Why is the line profile of the [CII] emission different from that of the CO and [CI]?''
Yuji Ebisawa pointed out that
we need to connect atomic cloud and molecular cloud, physically and chemically (CH/OH?).
Nadia Murillo is concerned on
characterization of the chemical structure of protostars and what factors influence it
She wishes to know if all protostars evolve the same chemically.
Naomi Hirano
wishes to know the origin of collimated molecular jet and its time scale.
Cecile Favre said
``There is evidence that the young Sun emitted a high flux of energetic ($>\sim$10 MeV) particles. This finding leads one to ask, whether and when this happened and how it affected the solar system and, do other protostars experience the same process now?''
Gwendoline St\'{e}phan said that
more laboratory data on photo-desorption are needed.
J.~S. Zhang wishes
to confirm the possible Galactic radial gradient, which needs a large sample of representative sources with accurate distance.

\subsection{Conclusion}
Still missing in understanding are 
UV-photons and cosmic-rays interactions with icy dust grains,
Sulphur and Phosphorus chemistry,
Metallicity effect,
Duration (time) effect,
COM formation and destruction,
phase transition (what we are seeing),
dust-gas interface,
dust evolution, etc.

What we should do are
multi-scale high spectral resolution molecular observations, laboratory work, theory, radiative transfer, etc.
We need careful modeling without simplifying things.

\section{Introduction to Research on Planck Cold Clumps}

\begin{figure*}[ht!]
\vskip -1.5cm
\includegraphics[angle=0,scale=0.6]{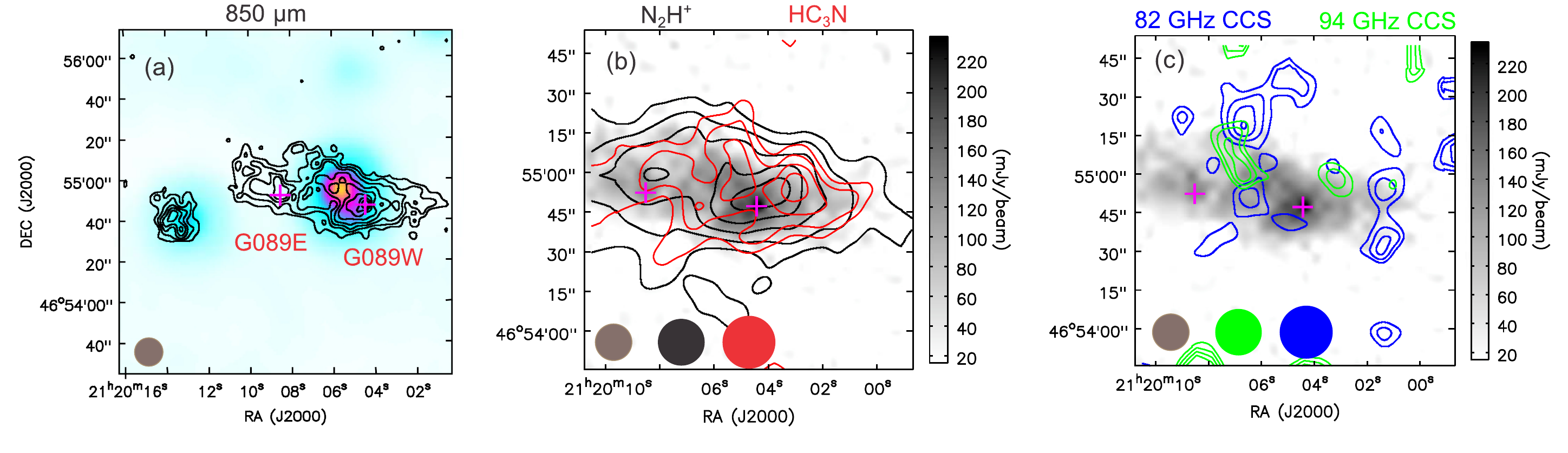}
\caption{(a) The 850 $\mu$m continuum image (contours) obtained toward G089.9-01.9 superimposed on the WISE 22 $\mu$m image (color).
The magenta cross symbol represents the SCUBA-2 peak position (Table 3).
The dark brown filled circle represents the half-power beam size for  JCMT/SCUBA-2 850 $\mu$m.
(b)
Black and red contour maps represent N$_2$H$^+$ and HC$_3$N integrated intensity maps, respectively, while
the gray-scale map represents the 850 $\mu$m continuum intensity.
The N$_2$H$^+$  integrated intensity is calculated for the main hyperfine component group $F_1$ = 2$-$1 ($F$ =  1$-$0,  2$-$1, and 3$-$2).
The black and red filled circles represent the half-power beam sizes for  N$_2$H$^+$ and HC$_3$N, respectively.
(c) Blue and green contours represent the 82 GHZ CCS and 94 GHz CCS integrated intensity maps, respectively, while 
the gray-scale map represents the 850 $\mu$m continuum intensity.
The blue and green filled circles represent the half-power beam sizes for  82 GHz CCS and 94 GHz CCS, respectively.
}
\end{figure*}

\subsection{Introduction} \label{sec:intro}

On the basis of the Planck all-sky survey \cite{2011AA...536A..23P,2016AA...594A..28P}, we are carrying out a series
of observations of molecular clouds as the Planck Cold Clump collaboration in order to
understand the initial condition for star formation \cite{2015PKAS...30...79L}. Planck cold clumps have low dust
temperatures (10$-$20 K; median=14.5 K). Pilot observations have been carried out with various ground-based telescopes such as
JCMT, IRAM, PMO 14m, APEX, Mopra, Effelsberg, CSO, and SMA \cite{2015PKAS...30...79L}.
A Large Program for JCMT dust continuum observations with SCUBA-2 (SCOPE
\footnote{https://www.eaobservatory.org/jcmt/science/large-programs/scope/})
and a Key Science Program with TRAO 14 m radio telescope (TOP
\footnote{http://radio.kasi.re.kr/trao/key\_science.php})
are ongoing. 

To characterize Planck cold clumps, it is essential to investigate their chemical and  physical properties
in detail. In particular, we try to make their evolutionary stages clear. The chemical
evolution of molecular clouds has been established to some extent, not only for nearby dark clouds
(e.g., \cite{1992ApJ...392..551S,1992ApJ...394..539H,1998ApJ...506..743B,2006ApJ...646..258H,2009ApJ...699..585H}, but also for giant
molecular clouds (GMCs;  Orion A GMC \cite{2010PASJ...62.1473T,2014PASJ...66...16T,2014PASJ...66..119O};  Vela C GMC \cite{2016PASJ...68....3O}; 
Infrared Dark Clouds \cite{2012ApJ...756...60S,2013ApJ...777..157H}). Carbon-chain molecules such as CCS and HC$_3$N tend to be abundant in starless molecular cloud cores, while N-bearing molecules such as NH$_3$ and N$_2$H$^+$ as well as c-C$_3$H$_2$ tend to be abundant in
star-forming molecular cloud cores.
Furthermore, deuterium fractionation ratios are powerful evolutionary tracers
\cite{2006ApJ...646..258H,2012ApJ...747..140S}.
We investigate the evolutionary stages of Planck cold clumps using molecular
column density ratios.
For this purpose, by using the Nobeyama 45 m telescope, we observed 13 Planck cold clumps, for
which we have already obtained accurate positions from preliminary IRAM 30m observations in N$_2$H$^+$ and/or SCUBA-2 observations.

\subsection{Observations}


Observations with the 15 m James Clerk Maxwell Telescope (JCMT) on Mauna Kea were made between 2014 November and 2015 December in the pilot survey phase (project IDs: M15AI05, M15BI061) 
of the JCMT legacy survey program ``SCOPE".  SCUBA-2 was employed for observations of the 850 $\mu$m continuum. It is a 10,000 pixel bolometer camera operating simultaneously at 450 and 850 $\mu$m.  The beam size of SCUBA-2 at 850 $\mu$m is $\sim$14". 
The typical rms noise level of the maps is about 6-10 mJy~beam$^{-1}$ in the central 3' area, and increases to 10-30 mJy~beam$^{-1}$
out to 12'.  The data were reduced using SMURF in the STARLINK package.

\begin{figure*}
\includegraphics[angle=0,scale=0.6]{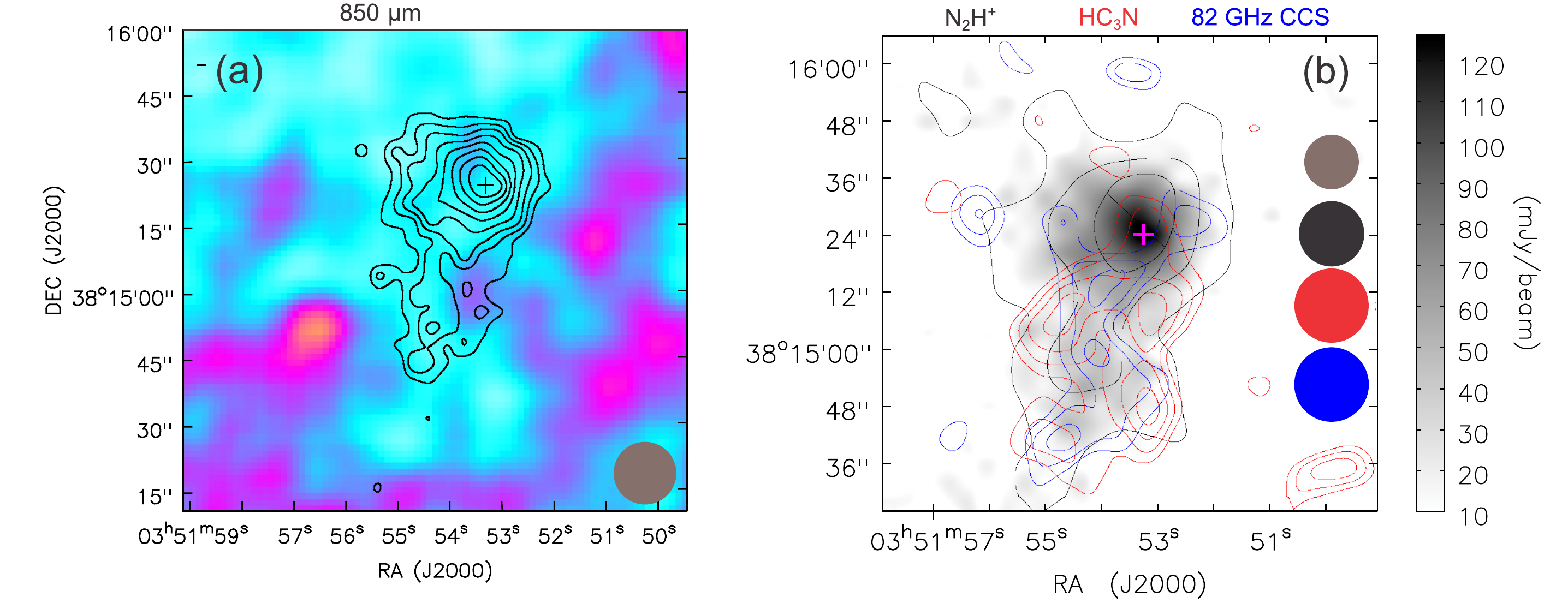}
\caption{Same as Figure 1 but for G157.6-12.2.
}
\end{figure*}

Observations with the 45 m radio telescope
of Nobeyama Radio Observatory\footnote{Nobeyama Radio Observatory
is a branch of the National Astronomical Observatory of Japan,
National Institutes of Natural Sciences.} were carried out from 2015 December to  2016 February.
Observations with the receiver TZ1 (we used one beam called TZ1 out of two beams of the receiver TZ) \cite{2013PASP..125..213A,2013PASP..125..252N} were made 
to simultaneously observe four molecular lines,
82 GHz CCS, 94 GHz CCS, HC$_3$N and N$_2$H$^+$.
Observations were carried out with T70 to simultaneously observe four other lines,
HN$^{13}$C, DNC, N$_2$D$^+$, and cyclic C$_3$H$_2$.
TZ1 and T70 are double-polarization, two-sideband SIS receivers.
The FWHM beam sizes at 86 GHz with TZ1 and T70 were 18".2$\pm$0".1 and 18".8$\pm$0".3, respectively.
The receiver backend was the digital spectrometer ``SAM45''.
The observed intensity is reported in terms of the corrected
antenna temperature $T_A^*$.
The observed data were reduced using the software package ``NoStar'' and ``NewStar''
of Nobeyama Radio Observatory.




\subsection{Results}

N$_2$H$^+$ was detected for all sources,
and we detected 81 GHz CCS from seven out of 13.
We show examples of  the molecular line intensity distribution shown in Figures 1 and 2.
In general, the N$_2$H$^+$ distribution (black contours in panel (b)) is quite similar to the 850 $\mu$m dust continuum emission distribution (contours in panel (a); gray-scale in panels (b) and (c)).
The 82 GHz CCS emission (blue contours) is clumpy in general, and is often located at the edge of the N$_2$H$^+$/850 $\mu$m  core or is distributed as it surrounds the N$_2$H$^+$/850 $\mu$m  core.
Most clumps are as cold as 10$-$20 K, and therefore the depletion of CCS can contribute to a configuration of the N$_2$H$^+$ core being surrounded by CCS \cite{2001ApJ...552..639A,2002ApJ...570L.101B}.

\subsection{Discussion}

\begin{figure}
\vskip -5.0cm
\hspace{-2cm}
\includegraphics[angle=0,scale=0.57]{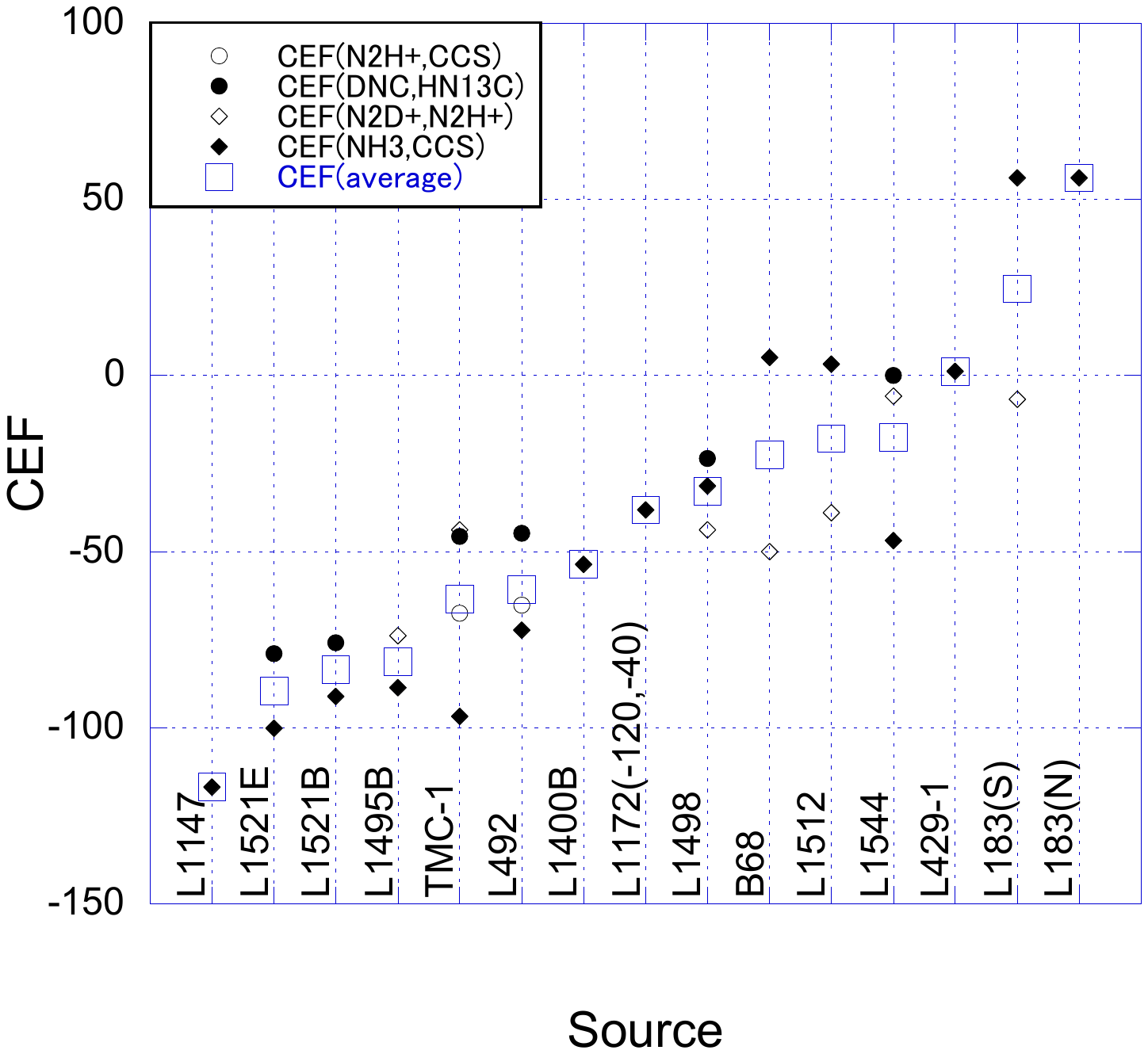}
\vskip -4.cm
\caption{Chemical Evolution Factor (CEF) for starless
sources in the literature.
}
\end{figure}

Hirota and Yamamoto\cite{2006ApJ...646..258H} indicated the evolutionary sequence of starless cores by using column density ratios such as $N$(DNC)/$N$(HN$^{13}$C).
We introduce a new parameter to represent the chemical evolution by using molecular column density ratios,
the chemical evolution factor (CEF).
We define CEF so that starless cores have CEFs of $\sim$ -100 to 0, 
and star-forming cores show $\sim$ 0 to 100.
Starless cores having CEF $\sim$ 0 are regarded as being on the verge of star formation.
By taking into account the observational results
\cite{1992ApJ...392..551S,2005ApJ...619..379C,2006ApJ...646..258H,2014PASJ...66...16T},
we define CEF as CEF =
log($N$(N$_2$H$^+$)/$N$(CCS)/2.5)*50,
log($N$(DNC)/$N$(HN$^{13}$C)/3)*120,
log($N$(N$_2$D$^+$)/$N$(N$_2$H$^+$)/0.3)*50,
and
log($N$(NH$_3$)/$N$(CCS)/70)*70,
for  
starless cores with $T_k$ $\sim$ 10$-$ 20  K at a spatial resolution of order 0.015-0.05 pc (for 0.1-pc sized structure ``molecular cloud core'').
Figure 3 shows the resulting CEF using the data in the literature 
\cite{2005ApJ...619..379C,2006ApJ...646..258H,2009ApJ...699..585H}.
Figures 4 and 5 show the CEFs estimated in the present study.
To see the effect of very different spatial resolution (and probably very different volume density and very different beam-filling factor),
we show sources located beyond 1 kpc in parentheses.
In this paper we treat only starless cores for CEF, because evolution of star-forming cores has not well been characterized yet.

\begin{figure}
\vskip -5.0cm
\hspace{-2cm}
\includegraphics[angle=0,scale=0.57]{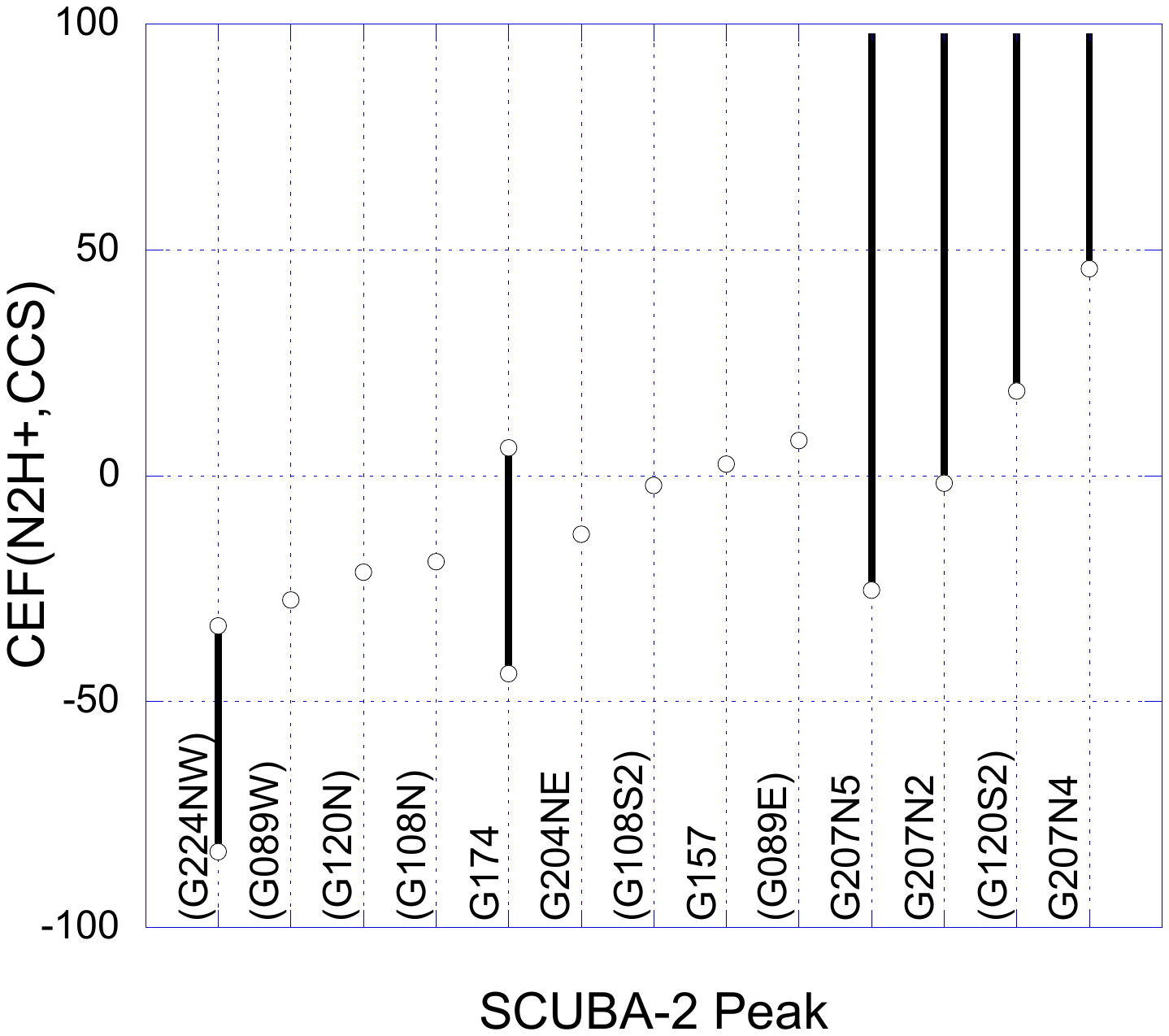}
\vskip -4.5cm
\caption{Chemical Evolution Factor (CEF) for starless SCUBA-2 peaks 
based on the column density ratio of $N$(N$_2$H$^+$)/$N$(CCS).
The source name in parentheses means distance is larger than 1kpc.
}
\end{figure}

\begin{figure}
\vskip -5cm
\hspace{-2cm}
\includegraphics[angle=0,scale=0.57]{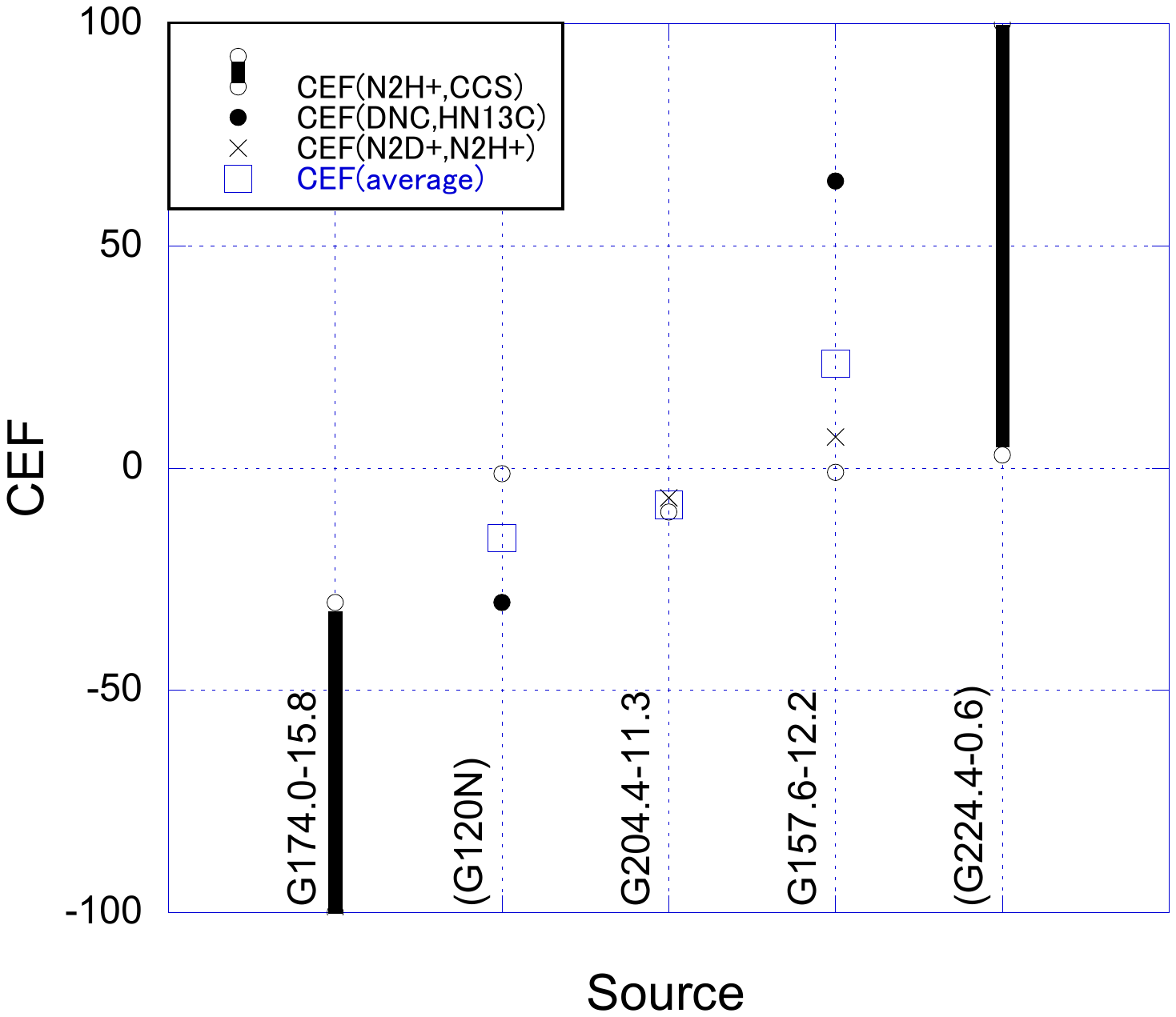}
\vskip -4.5cm
\caption{Chemical Evolution Factor (CEF) at starless T70 positions based on multiple column density ratios.
The source name in parentheses means distance is larger than 1kpc.
}
\end{figure}

Next, we investigate the morphology.  In G089.9-01.9, the 82 GHz and 94 GHz CCS emission (young molecular gas) is distributed as if it surrounds the N$_2$H$^+$ core (evolved gas). 
In G157.6-12.2, the 82 GHz CCS emission is distributed as if it surrounds the N$_2$H$^+$ core.
Such  configurations were previously reported in  L1544 \cite{2001ApJ...552..639A}, and also starless NH$_3$ core surrounded by CCS configurations are also observed in
L1498 \cite{2000ApJS..128..271L} and Orion A GMC \cite{2014ApJ...789...83T}. L1544 shows evidence of the prestellar
collapse.  
Therefore, these cores could be good targets for further studies for the initial conditions of star formation.
For  G157.6-12.2, CEF is $> \sim$ 0, and
its linewidth is as narrow as 0.3 km s$^{-1}$.
It is possible that this core is a coherent core that has largely dissipated turbulence, and is
on the verge of star formation 
\cite{2014ApJ...789...83T,2016MNRAS.459.4130O}.

The details are reported in
\cite{2017ApJS..228...12T}.

\section*{Acknowledgments}

I like to thank Fumitaka Nakamura (SOC Chair) and Quang Nguyen Luong (LOC Chair) for their great efforts for this wonderful conference.
The James Clerk Maxwell Telescope is operated by the East Asian Observatory 
on behalf of The National Astronomical Observatory of Japan, 
Academia Sinica Institute of Astronomy and Astrophysics, 
the Korea Astronomy and Space Science Institute, 
the National Astronomical Observatories of China and the Chinese Academy of Sciences (Grant No. XDB09000000), 
with additional funding support from the Science and Technology Facilities Council of the United Kingdom and participating universities in the United Kingdom and Canada.

\end{document}